# Gamma Ray Bursts

Neil Gehrels[1*] and Péter Mészáros[2]

[1] Astrophysics Science Division, NASA Goddard Space Flight Center, Greenbelt, MD 20771 USA

[2] Astronomy and Astrophysics Department, Pennsylvania State University, University Park, PA 16802 USA

*To whom correspondence should be addressed. E-mail: neil.gehrels@nasa.gov

Abstract --
Gamma-ray bursts (GRBs) are bright flashes of gamma-rays coming from the cosmos. They occur roughly once per day, last typically 10s of seconds and are the most luminous events in the universe. More than three decades after their discovery, and after pioneering advances from space and ground experiments, they still remain mysterious. The launch of the Swift and Fermi satellites in 2004 and 2008 brought in a trove of qualitatively new data. In this review we survey the interplay between these recent observations and the theoretical models of the prompt GRB emission and the subsequent afterglows.

GRBs are the most extreme explosive events in the Universe. The initial prompt phase lasts typically less than 100 s and has an energy content of $\sim 10^{51}$ ergs, giving a luminosity that is a million times larger than the peak electromagnetic luminosity of the bright emission from the exploding-star supernova phenomenon. The GRB name is a good one because their spectra peak in the gamma-ray band between ~100 keV and ~1 MeV. The source of the energy powering the bursts is thought to be the gravitational collapse of matter to form a black hole or other compact object.

GRBs were discovered in the late 1960s by the Vela satellites monitoring the Nuclear Test Ban Treaty between the US and the Soviet Union (1). It was slow progress for 20 years learning the origin of these brilliant flashes. Gamma-ray instruments of the time had poor positioning capability so only wide-field, insensitive telescopes could follow-up the bursts to look for counterparts at other wavelengths. Nothing associated with the GRBs was seen in searches hours to days after their occurrence. In the 1990's the Burst and Transient Source Experiment (BATSE) onboard the Compton Gamma Ray Observation (CGRO) obtained the positions of ~3000 GRBs and showed that they were uniformly distributed on the sky (2), indicating either an extragalactic or galactic-halo origin. BATSE also found that GRBs separate into two duration classes, short and long GRBs, with dividing line at ~2s (3). The detection of X-ray afterglows and more accurate localizations delivered by the BeppoSAX mission (4,5) enabled optical redshifts of GRBs to be measured and their extragalactic origin to be confirmed. The long bursts were found to be associated with far-away galaxies at typical redshifts z=1-2, implying energy releases in excess of $10^{50}$ ergs. The afterglow time decay often steepened after a day indicating a geometrical beaming of the radiation into a jet of opening angle ~5° (6).

Key open issues prior to the Swift and Fermi era included: (1) the origin of short GRBs, (2) the nature of the high energy radiation from GRBs, (3) the redshift distribution of bursts and their usage for early universe studies, and (4) the physics of the jetted outflows.

Swift and Fermi, launched in 2004 and 2008 respectively, have opened a new era in GRB research. They are both NASA missions with major international partnerships and have different and complementary capabilities. Swift has a wide-field imaging camera in the hard X-ray band that detects the bursts at a rate of ~100 per year, providing positions with arcminute accuracy. The spacecraft then autonomously and rapidly (100 s) reorients itself for sensitive X-ray and UV/optical observations of the afterglow. Fermi has two wide-field instruments. One detects bursts in the gamma-ray band at a rate of ~300 per year, providing spectroscopy and positions with 10-degree accuracy. The other observes bursts in the largely-unexplored high-energy gamma-ray band at a rate of ~10 per year. Combined, the two missions are advancing our understanding of all aspects of GRBs, including the origin of short bursts, the nature of bursts coming from the explosion of early stars in the universe and the physics of the fireball outflows that produce the gamma-ray emission.

## Swift GRB Observations

### Mission & Statistics

The Swift mission (7) has three instruments: the Burst Alert Telescope (BAT; 8), the X-Ray Telescope (XRT; 9) and the UV Optical Telescope (UVOT; 10). The BAT detects bursts and locates them to ~2 arcminute accuracy. The position is then sent to the spacecraft to repoint the XRT and UVOT at the event. Positions are also rapidly sent to the ground so that ground telescopes can follow the afterglows. There are more than 50 such telescopes of all sizes that participate in these world-wide follow-up campaigns. Measurements of the redshift and studies of host galaxies are typically done with large ground-based telescopes, which receive immediate alerts from the spacecraft when GRBs are detected. Swift has, by far, detected the largest number of well-localized bursts with afterglow observations and redshift determinations. As of 1 April 2012, BAT has detected 669 GRBs (annual average rate of ~90 per year). Approximately 80% of the BAT-detected GRBs have rapid repointings (the remaining 20% have spacecraft constraints that prevent rapid slewing). Of those, virtually all long bursts observed promptly have detected X-ray afterglow. Short bursts are more likely to have negligible X-ray afterglow, fading rapidly below the XRT sensitivity limit. The fraction of rapid-pointing GRBs that have UVOT detection is ~35%. Combined with ground-based optical observations, about ~60% of Swift GRB have optical afterglow detection. There are 200 Swift GRBs with redshifts compared with 41 pre-Swift.

A key finding of Swift is that the afterglow lightcurve has structure. Fast decay is often seen in the first 1000 s after the burst, followed by a shallow decay and then re-steeping (11, 12). Steeping due to beaming is sometimes seen, but not in every case (13, 14)

### Long GRBs

Long GRBs (LGRBs) are associated with the brightest regions of galaxies where the most massive stars occur (15). LGRBs occur over a large redshift range from z=0.0085 (GRB 980425) to z>8 (GRB 090423 & GRB 090429B) (Fig. 1). The redshift distribution for Swift is shown in Figure 1. With few exceptions, LGRBs that occur near enough for supernova detection have accompanying Type Ib or Ic supernovae. The exceptions are cases of bursts that may be misclassified as long or that may have exceptionally weak supernovae. These facts support the growing evidence that long bursts are caused by "collapsars" where the central core of a massive star collapses to a compact object such as a black hole (16) or possibly a magnetar (17, 18).

LGRBs are extremely bright in both gamma-ray prompt emission and multiwavelength afterglow. The typical optical/infrared brightness for several high-redshift (z>5) GRBs found by Swift is ~18th magnitude at a few hours after the event, compared to ~28th magnitude for a Milky-Way-type galaxy at redshift z = 5. This bright emission makes GRBs unique tools for studying the high-redshift universe: GRB 090423 at z=8.2 is the source with largest spectroscopically determined redshift (19,20). Multiwavelength observations of this and other high redshift bursts are providing information about the universe at a time when it was only about a few percent of its current age, and shed light on the process of reionization in the early universe.

The chemical evolution of the universe can be studied with GRBs as illustrated in Figure 2. GRBs provide data to higher redshift than active galaxies. The metallicity bias of GRBs is currently contradictory. Plots like Figure 2 show GRBs in higher metallicity star forming regions of galaxies. However, comparing star forming regions, GRBs actually tend to favor ones with lower metallicity (21).

Another way that GRBs are contributing to our understanding of the high-redshift universe is in the determination of the star formation history (Fig. 3). LGRBs are the endpoints of the lives of massive stars and their rate is therefore approximately proportional to the star formation rate. They give information at high redshift where the rate is highly uncertain. There may be evolutionary biases, such as a dependence of LGRBs on the metallicity of host galaxies, so studies relating them to the star formation rate must include these factors into account (22, 23).

Theories for the origin of LGRBs predate Swift, but are supported by the new data. In them, a solar rest mass worth of gravitational energy is released in a very short time (seconds or less) in a small region of the order of tens of kilometers by a cataclysmic stellar event. The energy source is the collapse of the core of a massive star. Only a small fraction of this energy is converted into electromagnetic radiation, through the dissipation of the kinetic energy of a collimated relativistic outflow, a fireball with bulk Lorentz factors of $\Gamma \sim 300$, expanding out from the central engine powered by the gravitational accretion of surrounding matter into the collapsed core or black hole.

Short GRB

At the time of Swift's launch, the greatest mystery of GRB astronomy was the nature of short-duration, hard-spectrum bursts (SGRBs). Although more than 50 long GRBs had afterglow detections, no afterglow had been found for any short burst. In summer 2005 Swift and HETE-2, precisely located three short bursts for which afterglow observations were obtained leading to a breakthrough in our understanding of short bursts (24-30). As of 2012, BAT has detected 65 SGRBs, 80% of which have XRT detections, and 15 of which have redshifts.

In contrast to long bursts, the evidence is that SGRBs typically originate in host galaxies with a wide range of star formation properties, including low formation rate. Their host properties are substantially different than those of long bursts (31, 32) indicating a different origin. Also, nearby SGRBs show no evidence for simultaneous supernovae (33 and references therein), very different than long bursts. Taken together, these results support the interpretation that SGRBs arise from an old populations of stars and are due to mergers of compact binaries (i.e., double neutron star or neutron star - black hole) (33-35)

Measurements or constraining limits on beaming from light curve break searches have been hard to come by with the typically weak afterglow of SGRBs. With large uncertainties associated with small number statistics, the distribution of beaming angles for SGRBs appears to range from ~5° to >25° (36, 37), roughly consistent but perhaps somewhat larger than that of LGRBs. Swift observations also revealed long (~100 s) "tails" with softer spectra than the first episode following the prompt emission for about 25% of short bursts (38, 39). Swift localization of a short GRB helps narrow the search window for gravitational waves from that GRB (40). Detection of gravitational waves from a Swift GRB would lead to great scientific payoff for merger physics, progenitor types, and NS equations of state.

**Fermi GRB Observations**

Mission & Statistics

The Fermi instruments are the Gamma-ray Burst Monitor (GBM, 41) and the Large Area Telescope (LAT, 42). The GBM has scintillation detectors and covers the energy range from 8 keV to 40 MeV. It measures spectra of GRBs and determines their position to ~5° accuracy. The LAT is a pair conversion telescope covering the energy range from 20 MeV to 300 GeV. It measures spectra of sources and positions them to an accuracy of <1°. The GBM detects GRBs at a rate of ~250 per year, of which on average 20% are short bursts. The LAT detects bursts at a rate of ~8 per year.

LAT bursts have shown two common and interesting features: (1) delayed emission compared to lower energy bands and (2) lower last prompt emission compared to lower every bands. The four brightest LAT bursts are GRB 080916C (43), GRB 090510 (44, 45), GRB 090902B (46), and GRB 090926A (47). They have yielded hundreds of >100 MeV photons each, and together with the lower energy GBM observations, have given unprecedented broad-band spectra. In GRB 080916C, the GeV emission appears only in a second pulse, delayed by ~4s relative to the first pulse (Fig. 4). Such a delay is present also in short bursts, such as GRB 090510, where it is a fraction of a second. This soft-to-hard spectral evolution is clearly seen in all four of these bright LAT bursts, and to various degrees a similar behavior is seen in other weaker LAT bursts.

In some bursts, such as GRB 080916C and several others, the broad-band gamma-ray spectra consist of a simple Band-type broken power-law function in all time bins. In GRB 080916C the first pulse has a soft high energy index disappearing at GeV energies, while the second and subsequent pulses have harder high energy indices reaching into the multi-GeV range. In some other bursts, such as GRB090510 (44, 45) and GRB 090902B (46), a second hard spectral component extending above 10 GeV without any obvious break appears in addition to common Band spectral component dominant in the lower 8 keV-10 MeV band.

An exciting discovery, unanticipated by results from the Energetic Gamma-Ray Experiment Telescope (EGRET) instrument on CGRO, was the detection of high-energy emission from two short bursts (GRB 081024B (45) and GRB 090510 (48)). Their general behavior (including a GeV delay) is qualitatively similar to that of long bursts. The ratio of detection rates of short and long GRBs by LAT is is ~7% (2 out of 27), which is significantly smaller than the ~20% by GBM.

While the statistics on short GRBs are too small to draw firm conclusions, so far, the ratio of the LAT fluence to the GBM fluence is >100% for the short bursts as compared to ~ 5 – 60% for the long bursts. It is also noteworthy that, for both long and short GRBs, the ≥100 MeV emission lasts longer than the GBM emission in the <1 MeV range. The flux of the long-lived LAT emission decays as a power law with time, which is more reminiscent of the smooth temporal decay of the afterglow X-ray and optical fluxes rather than the variable temporal structure in the prompt keV– MeV flux. This similarity in the smooth temporal evolution of the fluxes in different wave bands has been detected most clearly in GRB 090510 (49, 50) although this burst also requires a separate prompt component to the LAT emission (51). This short burst was at z = 0.9, and was jointly observed by the LAT, GBM, BAT, XRT and UVOT.

The LAT detects only ~10% of the bursts triggered by the GBM which were in the common GBM-LAT field of view. This may be related to the fact that the LAT-detected GRBs, both long and short, are generally among the highest fluence bursts, as well as being among the intrinsically most energetic GRBs. For instance, GRB 080916C was at z = 4.35 and had an isotropic-equivalent energy of $E_{iso} \approx 8.8 \times 10^{54}$ ergs in gamma rays, the largest ever measured from any burst (43). The long LAT bursts GRB 090902B (46) at z = 1.82 had $E_{iso} \approx 3.6 \times 10^{54}$ ergs, while GRB 090926A (47) at z = 2.10 had $E_{iso} \approx 2.24 \times 10^{54}$ ergs. Even the short burst GRB 090510 at z = 0.903 produced, within the first 2 s, an $E_{iso} \approx 1.1 \times 10^{53}$ ergs (45).

**Theoretical Models and Interpretation**

The current interpretation of the spectacular phenomenon of GRBs, whether long or short, is that it is likely to be related to the formation of a black hole, the large energy output being supplied by the gravitational energy being liberated in the process. This energy is liberated in a very short time, extending from seconds down to milliseconds, initiated either by the collapse of the rotating core of a massive star (52) in the case of long GRBs, or, as it seems for short GRBs, by the merger of two compact stellar remnants, such as two neutron stars or a perhaps a neutron star and a binary black hole (33-35), resulting in the disruption of the neutron star. While the massive collapse scenario for long bursts is fairly well established, the compact merger scenario of short bursts is likely but in need of further study. In both scenarios, the eventual result would be a black hole, although in some cases this may go through a temporary magnetar phase (a massive neutron star with an ultra-high magnetic field).

It is thought that this newly formed central engine or black hole then leads to the emission of an extremely intense gamma-ray pulse as a result of being fed, for a short period of time, by the infall of the rotating gas debris left over from the collapsing core or the merger. Some of this infall is accreted, but a larger fraction will end up being ejected in a jet along the rotation axis. The rotating debris is likely to lead to extremely strong magnetic fields, which couple the debris to the rotating black hole and, like super-strong rubber bands, extract the rotational energy of the black hole and pump it into the jet, which becomes highly relativistic and collimated into a 5-10 degree angular extent.

The energy of this jet is expected to be initially mainly in the form of kinetic energy of its motion. As it moves out from the black hole neighborhood, the initially large particle density in it decreases until at the photospheric radius the photon mean free path becomes larger than the jet dimension and the photons trapped in the jet can escape freely. However, if the jet energy is still mainly bulk kinetic energy at the photosphere, this escaping radiation would not amount to much, unless a substantial fraction of the (directed) kinetic energy has been dissipated into random energy of charged particles and radiated. A simple way this can occur is if the kinetic energy is dissipated beyond the photosphere in shocks, either internal shocks within the jet itself (53), or external shocks (54), as the jet is decelerated by external matter it encounters. Charged electrons bouncing across these shocks are accelerated via the Fermi mechanism to a relativistic power law energy distribution, and can produce a non-thermal photon spectrum via synchrotron or inverse Compton radiation, which approximates the observed Band type broken power law spectra. The simple internal shock interpretation of the prompt MeV emission, however, has typically low radiative efficiency, and the observed spectra sometimes disagree with a straightforward synchrotron interpretation, which has motivated searches for alternative interpretations (see next section).

The external shock will be accompanied by a reverse shock, which is expected to produce a prompt optical emission (55, 56), at the time the deceleration begins. This has been detected, with robotic ground-based telescopes such as ROTSE and others (52). As the jet continues to be decelerated in the course of sweeping up more and more external matter, the bulk Lorentz factor of the external shock decreases and the resulting nonthermal radiation becomes a long lasting, fading X-ray, optical and radio afterglow (56), whose predicted detection allowed the first measurements of host galaxies and redshift distances (52). The external shock interpretation of the late afterglow has proven robust overall. However, debate continues about some of the more detailed features seen in the first few hours by Swift, such as the X-ray steep decays followed by flat plateaus and occasional large flares, various proposed interpretations remaining to be tested.

**Recent MeV-GeV Models Based on Fermi and Swift**

The challenge of interpreting the Fermi-LAT GeV observations is most simply addressed through more or less conventional forward shock leptonic (i.e. relying on accelerated electrons or e+/e- pairs) synchrotron models (57, 58). Such models provide a natural delay between an assumed prompt MeV emission (e.g., from internal shocks or other inner mechanisms) and the GeV emission from the external shock, which starts after a few seconds time delay. However, taking into account the constraints provided by the Swift MeV and X-ray observations, it is clear that at least during the prompt emission, there must be an interplay between the shorter lasting mechanism providing the MeV radiation and the mechanism, or emission region, responsible for the bulk of the longer lasting GeV radiation (59, 60), the interplay between the two involving a number of subtleties. Unaddressed in these studies is a specific model of the prompt emission, and the above mentioned radiative inefficiency and spectral problems in a simple internal shock assumption. A resolution of this problem is possible if the prompt MeV Band spectrum is due to an efficient dissipative photosphere (baryonic, in this case) with an internal shock upscattering the MeV photons at a lower efficiency, giving the delayed GeV spectrum (61). Alternatively, for a magnetically dominated outflow, where internal shocks may not occur, an efficient dissipative photospheric Band spectrum can be up-scattered by the external shock and produce the observed delayed GeV spectrum (62). A delayed GeV spectrum may also be expected in hadronic models, which assume the co-acceleration, along with the electrons, of protons which undergo electromagnetic cascades and synchrotron losses along with their secondaries (63, 64).

Both Swift and Fermi are healthy and adequately-funded by NASA and foreign partners. Neither has lifetime limits from expendable propulsion gasses or cryogens, and both have orbital lifetimes beyond 2025. Key issues remain to be answered by the missions such as (1) confirmation of short burst origin, (2) jet opening angle and measurements of lighcurve breaks, (3) metallicity dependence, (4) relation to star formation history, and (5) nature and interpretation of high energy spectral components. We look forward to many years more of GRB discovery.

**Figures**

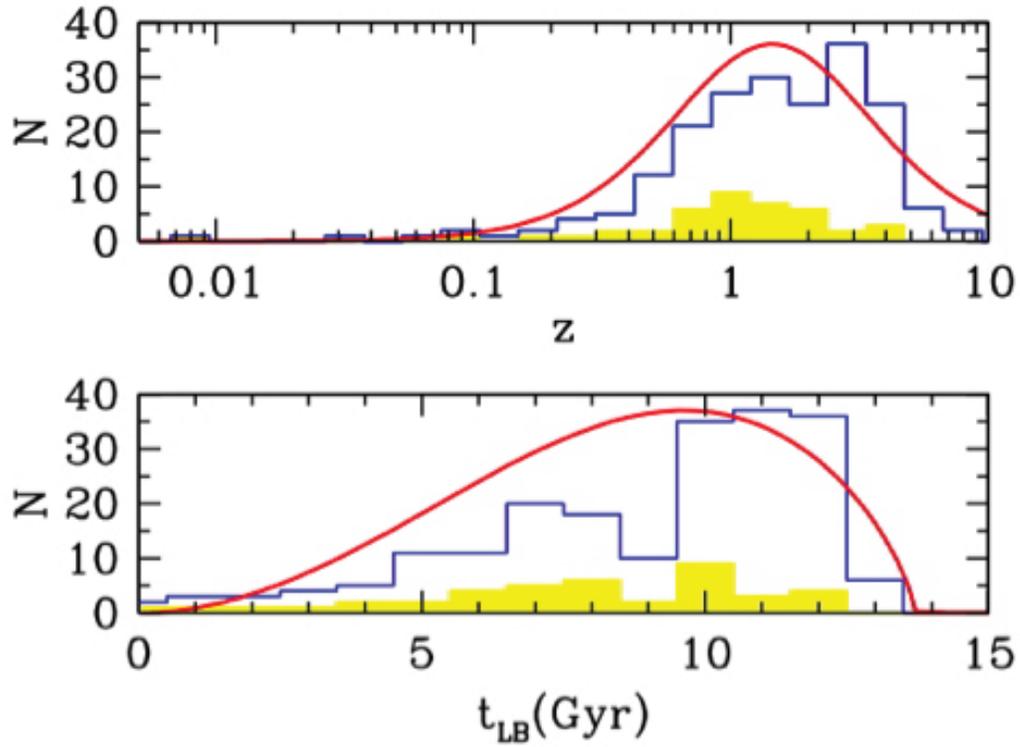

Figure 1 - Redshift distribution and cosmic look-back time of Swift long GRBs. The Swift GRBs are shown in blue, the pre-Swift GRBs in yellow and the co-moving volume of the universe with the red curve. The GRBs roughly follow the co-moving volume. From Gehrels, Ramirez-Ruiz and Fox (52)

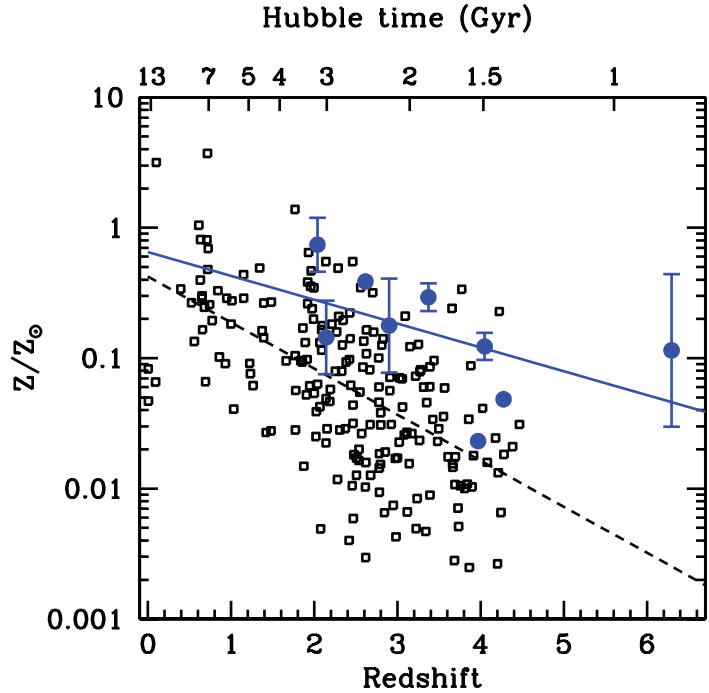

Figure 2 -Redshift evolution of the metallicity (represented here by the ratio of oxygen to hydrogen abundance) relative to solar values, for GRBs shown with blue dots and active galaxy quasars show with open circles. The abundances are determined from spectral absorption lines in the continuum radiation. GRB lines are predominantly from the gas in the host galaxy in the star-forming region near the explosion, whereas quasars are of random lines-of-sight through the galaxy. The GRB metallicity is on average ~5 times larger than in QSO. These are based on damped Lyman alpha (DLA) spectral features. The upper horizontal x-axis indicates the age of the Universe (Hubble time). From Savaglio et al. (65).

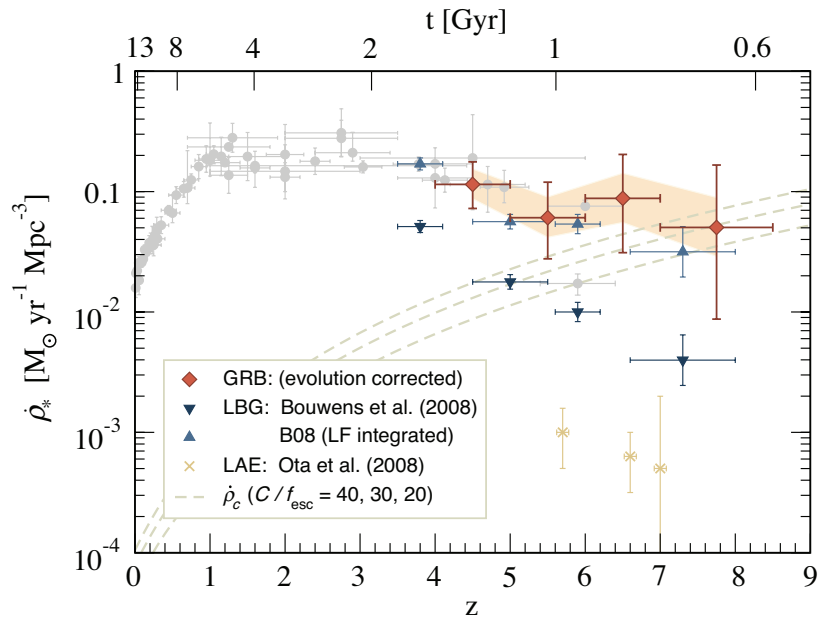

Figure 3 - From Kistler et al. (22). Cosmic star formation history. Shown are the data compiled in Hopkins & Beacom (66) (light circles) and contributions from Lyα emitters (LAE) (67). Recent LBG data are shown for two UV LF integrations: down to 0.2 L∗ (down triangles; as given in Bouwens et al. (68)) and complete z=3 (up triangles). Swift GRB-inferred rates are diamonds, with the shaded band showing the range of values resulting from varying the evolutionary parameters. Also shown is the critical density derivative from Madau et al. (69) for C/fesc = 40, 30, 20 (dashed lines, top to bottom). See also study by Robertson & Ellis (23)

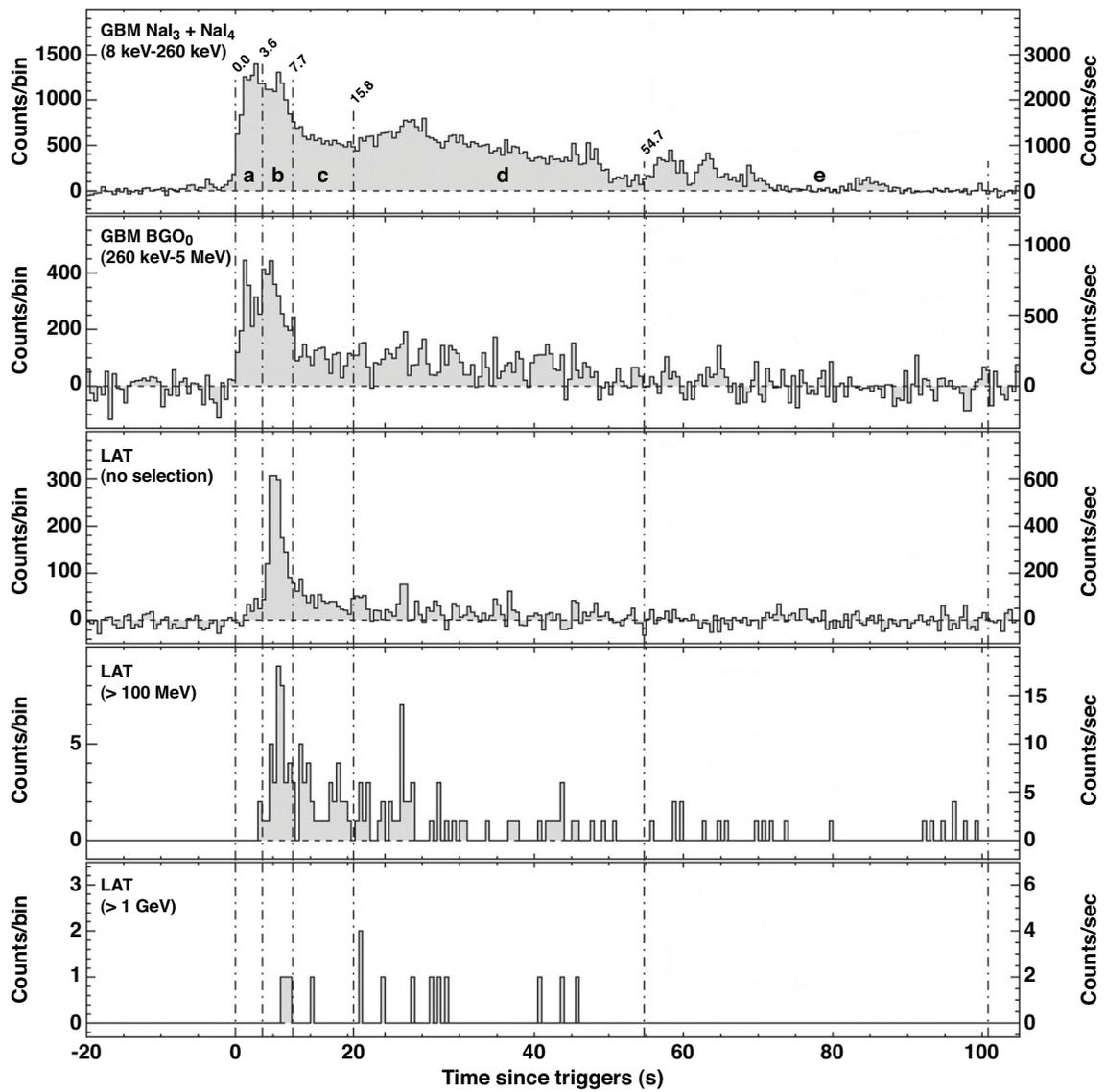

Figure 4 - Light curves of GRB 080916C with the GBM (top three panels) and LAT (bottom two panels). The high energy LAT emission is delayed relative to the lower energy GBM emission. From Abdo et al. (43).